\documentclass[english,aps,prb,reprint]{revtex4-1}
\usepackage[utf8]{inputenc}
\usepackage{geometry}
\usepackage{float}
\usepackage{amstext}
\usepackage{amssymb}
\usepackage{amsmath}
\usepackage{amsfonts}
\usepackage[hidelinks]{hyperref}
\usepackage{graphicx}
\usepackage{esint}
\usepackage[usenames, dvipsnames]{color}

\makeatletter
\@ifundefined{showcaptionsetup}{}{%
 \PassOptionsToPackage{caption=false}{subfig}}
\usepackage{subfig}
\makeatother

\usepackage{babel}
\begin{document}
\global\long\def\hc{\text{h.c.}}
\global\long\def\re{\text{Re}}
\global\long\def\im{\text{Im}}
\global\long\def\K{\mbox{ K}}
\global\long\def\meV{\mbox{ meV}}
\global\long\def\hf{\mbox{\ensuremath{_{2}F_{1}}}}
\global\long\def\pr{\prime}

\title{Power-law liquid in cuprate superconductors from fermionic unparticles}

\author{Zhidong Leong}
\author{Chandan Setty}
\author{Kridsanaphong Limtragool}
\author{Philip W. Phillips}

\affiliation{Department of Physics and Institute for Condensed Matter Theory,
University of Illinois, Urbana, Illinois 61801, U.S.A}

\date{\today}
\begin{abstract}
Recent photoemission spectroscopy measurements [arXiv:1509.01611] on cuprate superconductors
have inferred that over a wide range of doping, the imaginary part of the electron
self-energy scales as $\Sigma^{\prime\prime}\sim(\omega^2+\pi^2T^2)^a$ with $a=1$ in the overdoped Fermi-liquid state and 
$a<0.5$ in the optimal to underdoped regime. We show that this non-Fermi-liquid scaling behavior can naturally
be explained by the presence of a scale-invariant state of matter
known as unparticles. We evaluate analytically the electron self-energy
due to interactions with fermionic unparticles. We find that, in agreement
with experiments, the imaginary part of the self-energy scales with
respect to temperature and energy as $\Sigma^{\prime\prime}\sim T^{2+2\alpha}$
and $\omega^{2+2\alpha}$, where $\alpha$ is the anomalous dimension
of the unparticle propagator. In addition, the calculated occupancy
and susceptibility of fermionic unparticles, unlike those of normal
fermions, have significant spectral weights even at high energies.
This unconventional behavior is attributed to the branch cut
in the unparticle propagator which broadens the unparticle spectral
function over a wide energy range and non-trivially alters the scattering
phase space by enhancing (suppressing) the intrinsic susceptibility at low energies for negative (positive) $\alpha$. Our work presents new evidence suggesting that unparticles
might be important low-energy degrees of freedom in strongly coupled systems such as the
cuprate superconductors.
\end{abstract}
\maketitle

\section{Introduction}

Understanding the physics of cuprate superconductors involves identifying
the low-energy degrees of freedom that can reproduce the bizarre
features of the normal state which traditionally include $T$-linear resistivity, pseudogap, Fermi arcs, etc. Adding to the complexity are the 
recent angle-resolved photoemission spectroscopy (ARPES) measurements
\cite{Reber2015} of the cuprates that revealed that its well-known $T$-linear
resistivity can be construed as a slice of a unified power-law scaling
behavior. Over a wide range of doping levels, the measured scattering
rates in the non-superconducting state scale with respect to temperature
and frequency as $\Sigma^{\prime\prime}\sim\left(\omega^{2}+\pi^{2}T^{2}\right)^{a}$,
with only the scaling exponent $a$ varying with doping. The power-law
smoothly varies from Fermi-liquid-like at overdoping, to one with
$a\sim0.5$ representing $T$-linear scattering rate at optimal doping,
and to $a\lesssim0.5$ at underdoping. Such a non-Fermi-liquid state
of matter is dubbed a \textit{power-law liquid}.

Theoretically, mechanisms yielding similar non-Fermi liquid scalings
have been extensively studied \cite{Varma1989,Metzner2003,Lee2009a,Watanabe2014a,Fitzpatrick2014,Sachdev2011,Faulkner2010a,Casey2011,Faulkner2010b,Zaanen2015}.
In a marginal Fermi liquid \cite{Varma1989}, a polarizability proportional
to $\omega/T$ leads to $T$-linear resistivity, while a $d$-wave
Pomeranchuk instability in two dimensions \cite{Metzner2003} yields
self-energies with $\omega^{2/3}$ and $T^{2/3}$ dependence. In addition,
similar behaviors can also be obtained by coupling quasiparticles
with gauge bosons \cite{Lee2009a}, Goldstone bosons \cite{Watanabe2014a},
and critical bosons \cite{Fitzpatrick2014} near a quantum critical
point \cite{Sachdev2011}. Furthermore, strong coupling theories using the
anti-de Sitter spacetime (AdS)/conformal field theory (CFT) correspondence
\cite{Faulkner2010a} and Gutzwiller projection in hidden Fermi liquid
theory \cite{Casey2011} also exhibit $T$-linear resistivity. In
particular, the spectral functions calculated within the AdS/CFT formalism
can also exhibit a range of power-law scaling when the scaling dimension
of the boundary fermionic operator is tuned continuously \cite{Faulkner2010b,Zaanen2015}. 

Because of the recent unified scaling observations, it is natural to invoke
a scale-invariant sector such as unparticles as the effective 
low-energy degrees of freedom in the cuprates. Proposed a decade ago
as a scale-invariant sector within the standard model \cite{Georgi2007},
unparticles can emerge in strong coupling theories as low-energy degrees
of freedom. Exhibiting features similar to those of a fractional number of
invisible massless particles \cite{Georgi2007}, unparticles are an
incoherent state of matter that lack any particle-like behavior. They
can be construed as a product of states with a continuous distribution
of masses \cite{Stephanov2007,Krasikov2007,Deshpande2008} and can be constructed from theories
in AdS \cite{Cacciapaglia2009}.

While extensively studied in high-energy physics, unparticles remain
relatively new in condensed matter physics. In the context of the
cuprates, unparticles have been proposed to explain the absence of
Luttinger's theorem in the pseudogap phase \cite{Phillips2013} using
zeros in the Green function \cite{Dave2013} and have also been
found to yield unusual superconducting properties \cite{Phillips2013,LeBlanc2015,Karch2016}
and optical conductivity \cite{Limtragool2015}. 

In this paper, we show analytically that interactions between electrons
and fermionic unparticles can reproduce the power-law liquid revealed
in the cuprates by recent ARPES experiments \cite{Reber2015}. This
paper is a follow-up to our recent paper that focused on bosonic unparticles
\cite{Limtragool2016}. Here we find that, in agreement with the experiments,
the electron self-energy due to interactions with fermionic unparticles
exhibits power-law scaling with respect to both energy and temperature:
$\Sigma^{\prime\prime}\sim\omega^{2+2\alpha}$ and $T^{2+2\alpha}$,
where $\alpha$ is the anomalous scaling of the unparticle propagator.
In addition, we find that the occupancy number and susceptibility
of fermionic unparticles, unlike those of normal fermions, have significant
spectral weights even at high energies. These unconventional behaviors
can be attributed to the branch cut in the unparticle propagator
which  broadens the unparticle spectral function over a wide energy
range, and non-trivially alters the scattering phase space by enhancing (suppressing) the intrinsic susceptibility at low energies for negative (positive) $\alpha$.

\section{Electron-Fermionic Unparticle Scattering}
\subsection{Model}

We consider a system of electrons in the presence of a background of fermionic unparticles. The action of the system in Matsubara-Fourier space is given by 
\begin{align} \label{eq:action}
S = & \ T\sum\limits_{n}\sum\limits_{p} \psi_{n}^\dagger(p) G^{-1}_0(p,i\omega_n) \psi_{n}(p) \nonumber \\
 & + T\sum\limits_{n}\sum\limits_{p}\phi_{n}^\dagger(p) G^{-1}_\alpha(p,i\omega_n) \phi_{n}(p) \nonumber \\
 & + U T^3\sum\limits_{m,n,l}\sum\limits_{k,p,q} \psi_{m-l}^\dagger(k-q)\phi_{n+l}^\dagger(p+q)\phi_{n}(p)\psi_{m}(k),
\end{align}
where $\psi$ is the non-relativistic electron field, $\phi$ is the fermionic unparticle field, $G_0$ is the bare electron Green function 
\begin{equation}
G_0(p,i\omega_n) = \frac{1}{i\omega_n - E_p},
\end{equation}
and $G_\alpha$ is the fermionic unparticle Green function 
\begin{eqnarray}
G_{\alpha}\left(k,i\omega_{n}\right) & = & \frac{1}{\left(i\omega_{n}-\epsilon_{k}+\mu\right)^{1-\alpha}}.\label{eq:Greens-func}
\end{eqnarray}
Here, $\epsilon_{k}$ is the unparticle
energy spectrum, $1-\alpha$ is the scaling exponent, and $\mu$ is the chemical potential. When $\alpha=0$,
the Green function reduces to that of a normal particle. 
In addition, $U$ is the interaction between electrons and unparticles, and $T$ is the temperature. The subscripts of the fields denote the dependence on the Matsubara frequency. In this model, the fermionic unparticles are assumed to exist up to a UV momentum cutoff, $\Lambda$ because they represent a low-energy description of some microscopic theory.
For the
unparticle Green function to be scale-invariant, we set $\mu=0$
when $\alpha\neq0$. While the literature in high-energy physics considers
fermionic unparticles as relativistic four-spinors within the standard
model \cite{Luo2008,Basu2009}, here in the context of the cuprates,
we consider them as non-relativistic fermions. For simplicity, we also omit the normalization factor and the effects of spins. 

\begin{figure}[H]
\begin{centering}
\subfloat[]{\begin{centering}
\includegraphics[width=0.8\columnwidth]{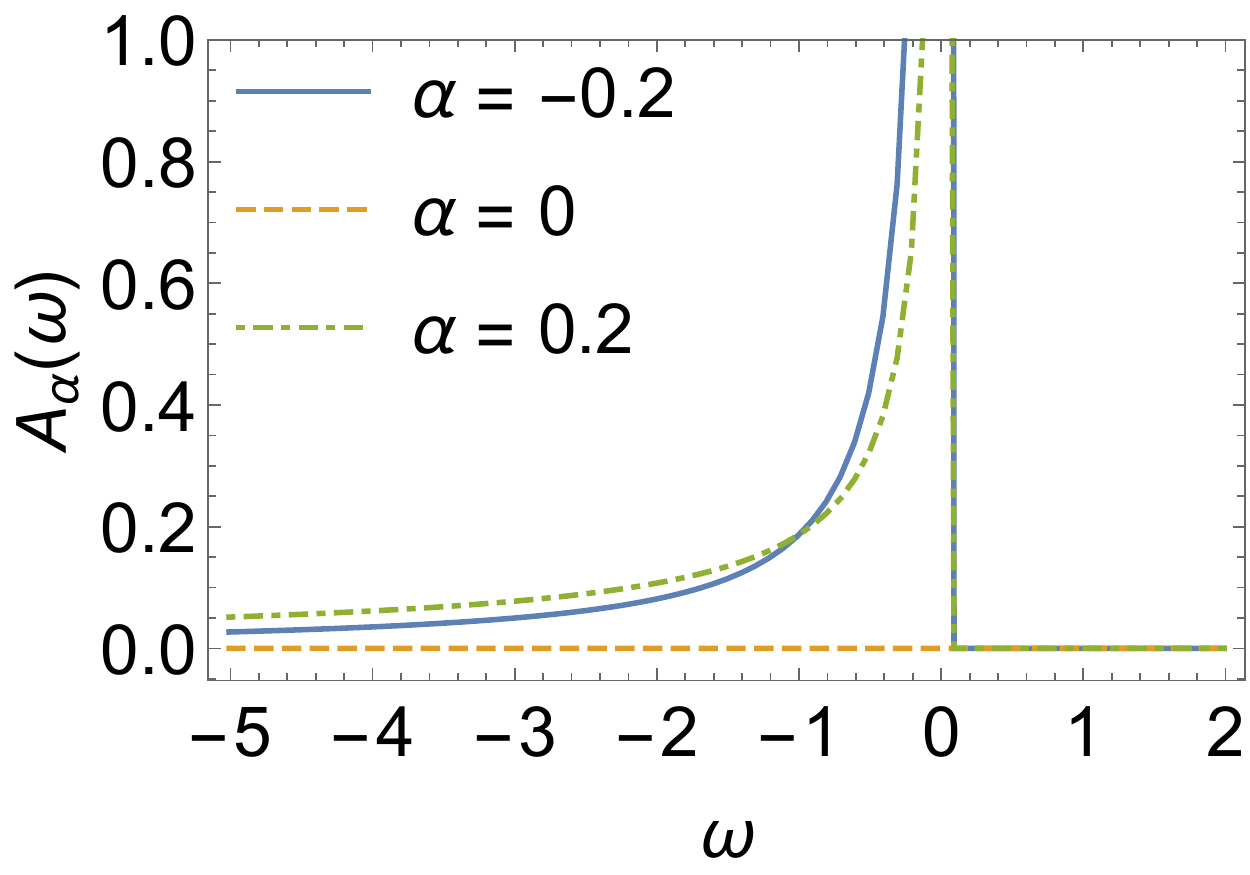}
\par\end{centering}
}
\par\end{centering}
\begin{centering}
\subfloat[]{\begin{centering}
\includegraphics[width=0.7\columnwidth]{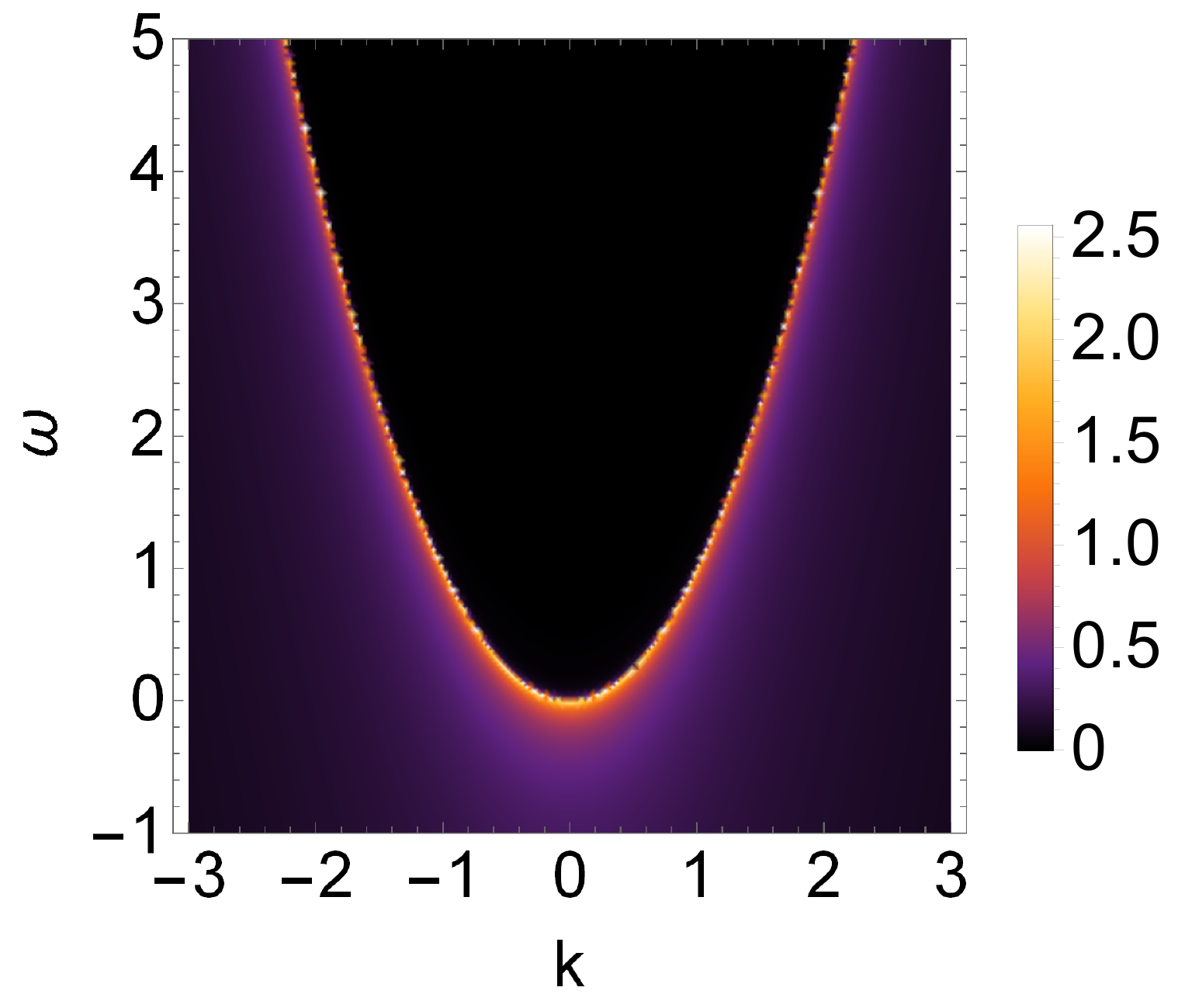}
\par\end{centering}
}
\par\end{centering}
\caption{(a) The spectral function $A_{\alpha}\left(\omega\right)$
of unparticles compared to that of particles. As $\alpha$ deviates
from zero, the delta peak in the spectral function broadens due to
the branch cut in the Green function. (b) The energy and momentum
dependence of the unparticle spectral function for a quadratic energy
spectrum $\epsilon_{k}\sim k^{2}$ with $\alpha=0.5$. The broadening
of the spectral function reflects the incoherent nature of unparticles. 
\label{fig:Spectral-function-plot}}
\end{figure}

In this paper, we focus on unparticles with $-1<\alpha<1$.
In this case, instead of a simple pole, the unparticle Green function
has a branch cut, which we choose to be along the negative energy
axis. That is, the branch cut of $z^{1-\alpha}$ is chosen to be along
$-\infty<z<0$ with the phase angle defined in the range $-\pi<\theta<\pi$. Fig. \ref{fig:Spectral-function-plot} shows that,
compared to particles, the spectral function of unparticles
\begin{eqnarray}
A_{\alpha}\left(k,\omega\right) &\equiv& -\frac{1}{\pi}\mathrm{Im} G_\alpha(k,\omega+i\eta) \nonumber \\
&=&\frac{1}{\pi}\left|\sin\left(\pi\alpha\right)\right|\frac{\theta\left(\epsilon_{k}-\omega\right)}{\left|\epsilon_{k}-\omega\right|^{1-\alpha}}\label{eq:spectral_func}
\end{eqnarray}
remains divergent at $\omega=\epsilon_{k}$, but has a broadened peak
due to the presence of the branch cut, representing the incoherence
of unparticles. Here $\theta(x)$ is the Heaviside step function.  It is precisely the modeling of the broad incoherent background in the electron spectral function that unparticles are tailored to handle.

\subsection{Electron self-energy}

\begin{figure}[H]
\begin{centering}
\includegraphics[width=0.45\columnwidth]{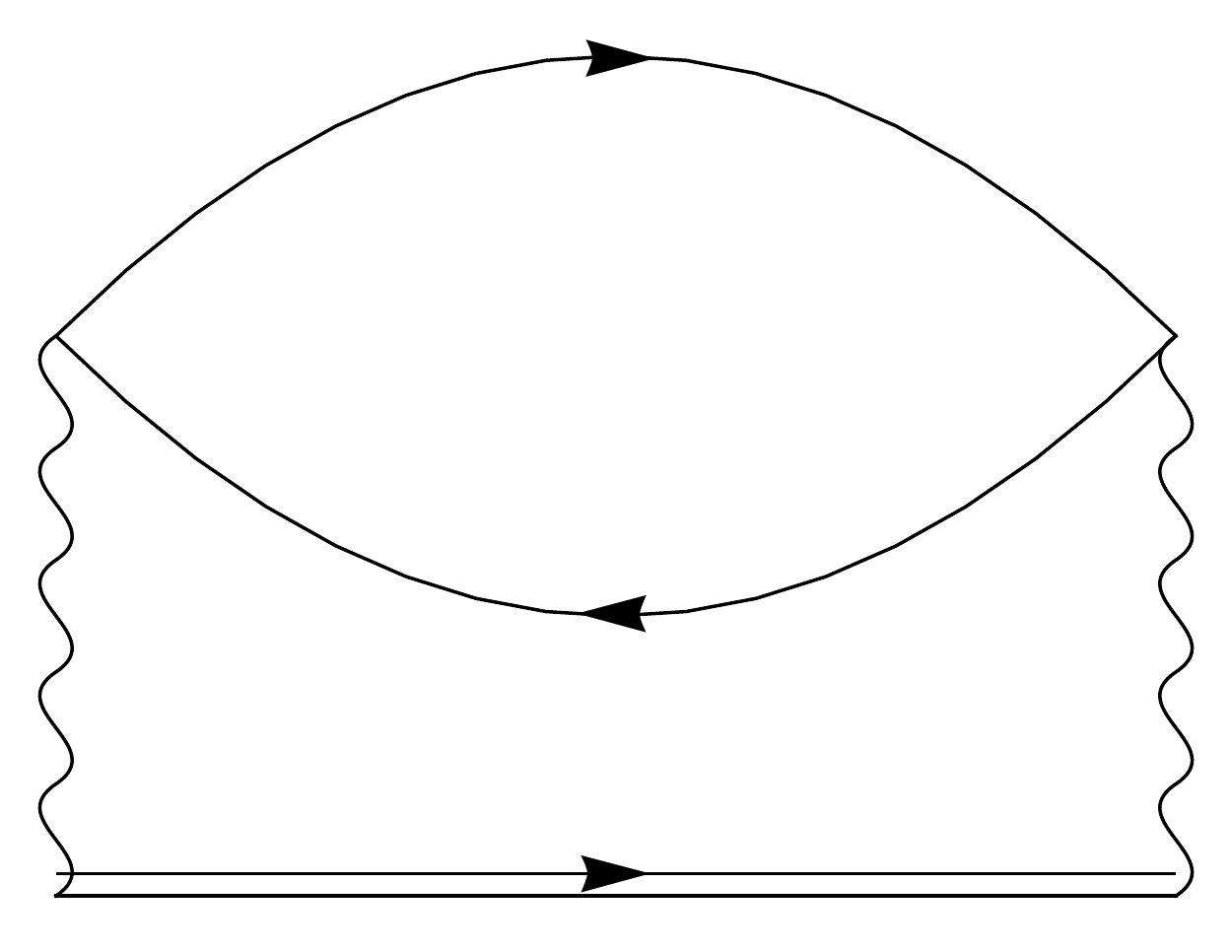}
\par\end{centering}
\caption{The lowest-order Feynman diagram of the electron self-energy due to
interactions between electrons and fermionic unparticles. The solid
lines, double line, and wavy lines correspond to fermionic unparticles,
electron, and the electron-unparticle interaction, respectively. \label{fig:self-energy_feynman-diagram}}
\end{figure}

For a constant interaction $U$ between electrons and fermionic unparticles,
Fig. \ref{fig:self-energy_feynman-diagram} illustrates the lowest-order
contribution to the electron self-energy $\Sigma\left(k,i\omega_{n}\right)$.
This can be written as 
\begin{eqnarray}
\Sigma\left(k,i\omega_{n}\right) & = & -U^2 \sum_{q}T\sum_{i\omega_{m}}G_{0}\left(k-q,i\omega_{n}-i\omega_{m}\right)\nonumber \\
&  & \qquad\qquad\qquad\qquad \chi_{\alpha}\left(q,i\omega_{m}\right),
\label{eq:self-energy}
\end{eqnarray}
where 

\begin{eqnarray}
\chi_{\alpha}\left(q,i\omega_{m}\right) & = & \sum_{p}T\sum_{i\omega_{n}}G_{\alpha}\left(p,i\omega_{n}\right)G_{\alpha}\left(p-q,i\omega_{n}-i\omega_{m}\right)\nonumber \\
\label{eq:suscep}
\end{eqnarray}
is the unparticle susceptibility, and $G_{0}\left(p,i\omega_{m}\right)$
is the electron Green function. While unparticle-particle interactions
in the standard model are constrained by experiments to be weak
\cite{Georgi2007}, the coupling strength $U$ here in the cuprates
can be significant.

Appendix \ref{app:Matsubara-sums} details our analytic evaluation of
the Matsubara sums in Eqs. \ref{eq:self-energy} and \ref{eq:suscep}
using standard contour integration techniques. After analytic continuation
$i\omega_{n}\rightarrow\omega+i\eta$, we find that the imaginary part of the electron
self-energy can be written as the momentum sum
\begin{eqnarray}
\Sigma^{\prime\prime}\left(k,\omega\right) & = & -U^2\sum_{pq}\tilde{S}_{\alpha}^{\pr\pr}\left(\epsilon_{p},\epsilon_{p-q},E_{k-q},\omega\right),
\end{eqnarray}
where 

\begin{figure}[H]
\centering{}\includegraphics[width=0.9\columnwidth]{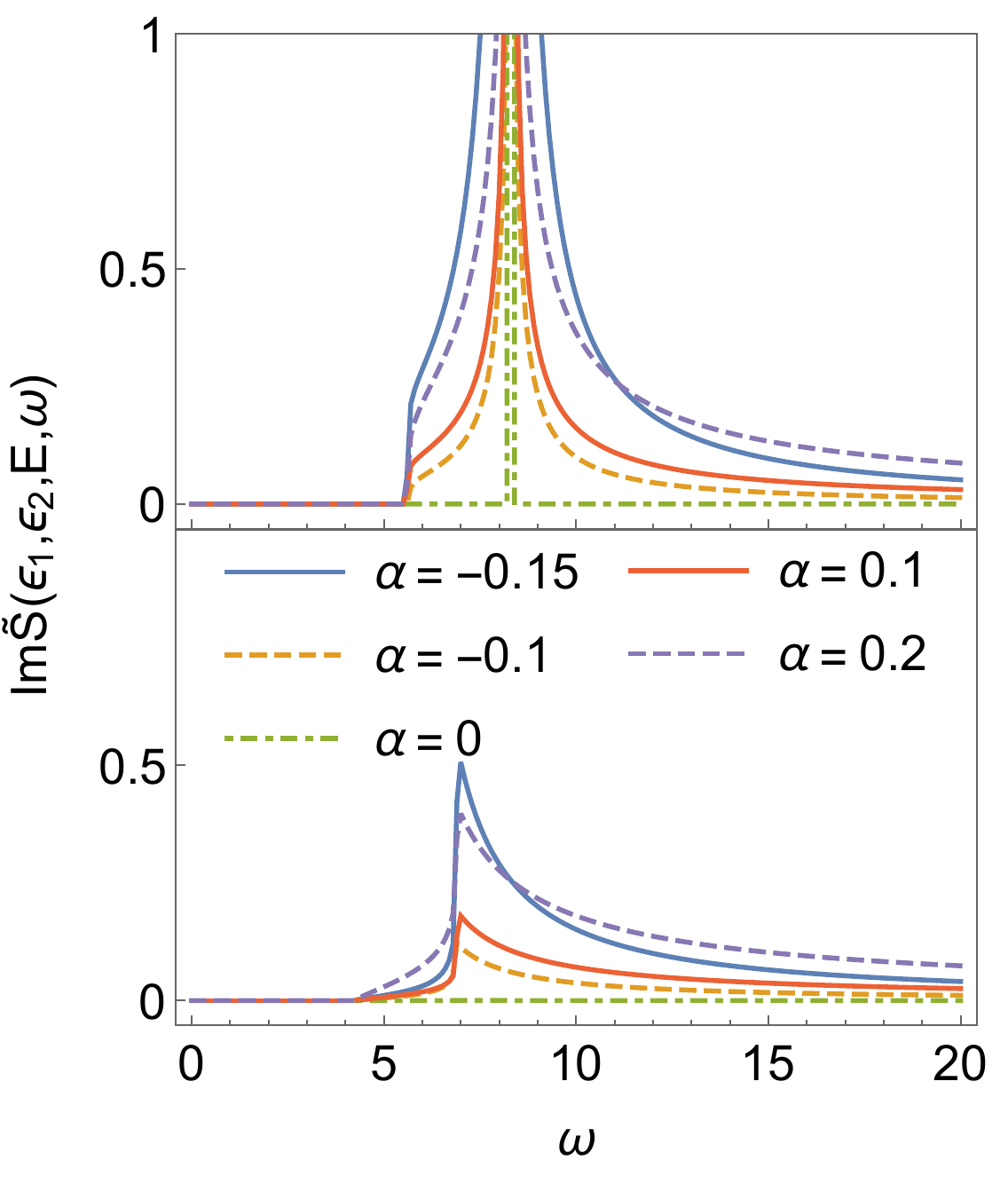}\caption{Top: Plot of the self-energy Matsubara sum $\tilde{S}_{\alpha}^{\protect\pr\protect\pr}\left(\epsilon_{1},\epsilon_{2},E,\omega\right)$
at $T=0.01$ for $\epsilon_{1}=2.7$, $\epsilon_{2}=-1.4$, and $E=4.2$.
Bottom: Same plot but with $\epsilon_{2}=1.4$. Compared to the Fermi
liquid result, the unparticle one has a broadened divergent peak and
additional nontrivial contributions. \label{fig:self-energy-Matsum-plot}}
\end{figure}

\begin{widetext}

\begin{eqnarray}
\tilde{S}_{\alpha}^{\pr\pr}\left(\epsilon_{1},\epsilon_{2},\epsilon_{3},\omega\right) & = & \left[n_{B}\left(\omega-\epsilon_{3}\right)+n_{F}\left(-\epsilon_{3}\right)\right]\left[\bar{\kappa}_{\alpha}\left(\epsilon_{2},\epsilon_{1}-\omega+\epsilon_{3}\right)-\bar{\kappa}_{\alpha}\left(\epsilon_{1},\epsilon_{2}+\omega-\epsilon_{3}\right)\right],\label{eq:self-energy-Matsum}\\
\bar{\kappa}_{\alpha}\left(\epsilon,\epsilon^{\pr}\right) & = & \pi\int_{-\infty}^{\infty}dz\ n_{F}\left(z\right)A_{\alpha}\left(z-\epsilon\right)A_{\alpha}\left(z-\epsilon^{\pr}\right).\label{eq:kappa-def}
\end{eqnarray}
\end{widetext}
Here, $n_{F/B}\left(z\right)=\left(e^{z/T}\pm1\right)^{-1}$ is the
Fermi (Bose) distribution, and $E_{k}$ is the electron energy spectrum.
To elucidate the analytic structure of $\tilde{S}_{\alpha}^{\pr\pr}$,
we note that, for $\alpha>0$ in the $T\rightarrow0$ limit, the integral
$\bar{\kappa}$ evaluates to the closed-form expression

\begin{eqnarray}
\bar{\kappa}\left(\epsilon,\epsilon^{\prime}\right) & = & \frac{1}{1-2\alpha}\frac{1}{\pi}\sin^{2}\left(\pi\alpha\right)\left[\frac{2}{\xi\left(\epsilon,\epsilon^{\prime}\right)}\right]^{1-2\alpha}\nonumber \\
 &  & \hf\left[1-\alpha,\frac{1}{2}-\alpha;\frac{3}{2}-\alpha;\left|\frac{\epsilon-\epsilon^{\prime}}{\xi\left(\epsilon,\epsilon^{\prime}\right)}\right|^{2}\right],
\end{eqnarray}
where $\hf\left(a,b;c;z\right)$ is the hypergeometric function, and
$\xi\left(\epsilon,\epsilon^{\prime}\right)=\max\left(\left|\epsilon-\epsilon^{\prime}\right|,\epsilon+\epsilon^{\prime}\right)$.
As in the Fermi liquid case, $\alpha=0$, 
\begin{eqnarray}
\tilde{S}_{\text{FL}}^{\pr\pr}\left(\epsilon_{1},\epsilon_{2},\epsilon_{3},\omega\right) & = & \pi\delta\left(\epsilon_{1}-\epsilon_{2}-\omega+\epsilon_{3}\right)\nonumber \\
 &  & \ \left[\theta\left(-\epsilon_{1}\right)\theta\left(-\epsilon_{3}\right)\theta\left(\epsilon_{2}\right)\right.\nonumber \\
 &  & \ \left.+\theta\left(\epsilon_{1}\right)\theta\left(\epsilon_{3}\right)\theta\left(-\epsilon_{2}\right)\right],\label{eq:FL-self-energy-Matsum}
\end{eqnarray}
we find that the analogous expression for unparticles $\tilde{S}_{\alpha}^{\pr\pr}\left(\epsilon_{1},\epsilon_{2},\epsilon_{3},\omega\right)$
diverges when $\epsilon_{1}-\epsilon_{2}+\epsilon_{3}-\omega=0$ and
$\epsilon_{1}\epsilon_{2}<0$. 
However, given that the unparticle chemical potential $\mu=0$, this divergence does not occur because $\epsilon_1,\epsilon_2$ are nonnegative. In addition, unlike the Fermi
liquid result, $\tilde{S}_{\alpha}^{\pr\pr}\left(\epsilon_{1},\epsilon_{2},\epsilon_{3},\omega\right)$
can be nonzero for other values of energies $\epsilon_{1},\epsilon_{2},\epsilon_{3},\omega$.
These features are illustrated in Fig. \ref{fig:self-energy-Matsum-plot}.
These nonzero values provide additional contributions to the electron
self-energy, and can be attributed to the broadening of the unparticle
spectral function illustrated in Fig. \ref{fig:Spectral-function-plot}.

\begin{figure}[H]
\begin{centering}
\includegraphics[width=0.9\columnwidth]{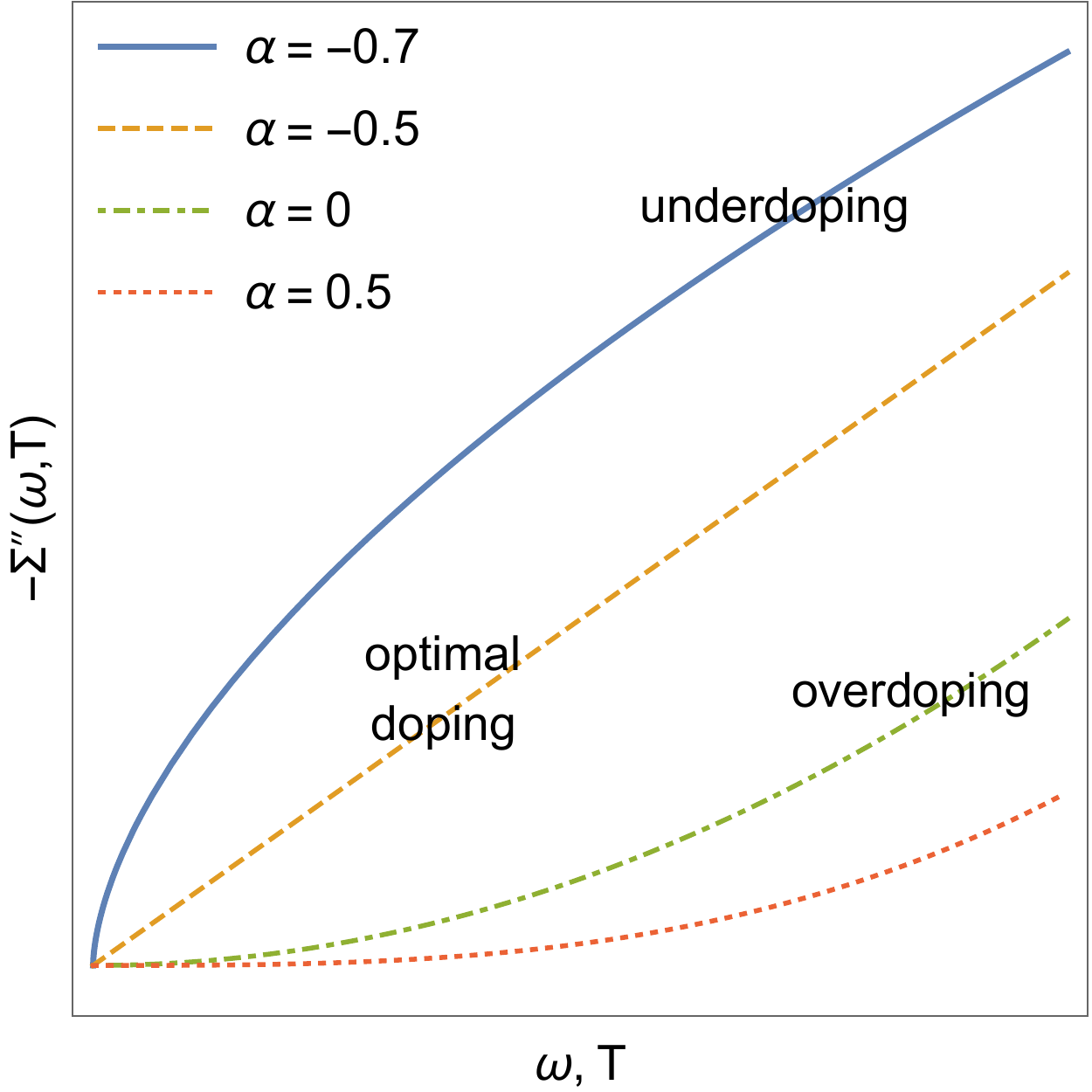}
\par\end{centering}
\caption{Schematic of the energy and temperature dependence of the electron
self-energy, showing deviations from Fermi liquid theory. In the cuprates,
unparticles with $\alpha\lesssim0$, $\alpha\approx-0.5$, and $\alpha<-0.5$
correspond to overdoping, optimal doping, and underdoping, respectively.
\label{fig:self-energy-plot}}
\end{figure}

Next, to determine the scaling form of the electron self-energy in
the $T\rightarrow0$ limit, we note that the unparticle spectral function scales as 
\begin{eqnarray}
A_\alpha(\lambda\omega) & = & -\frac{1}{\pi}\lim_{\eta\rightarrow 0}\im G_\alpha(\lambda\omega+i\eta) \nonumber \\
& = & -\frac{1}{\pi}\lambda^{-1+\alpha}\lim_{\eta\rightarrow 0}\im G_\alpha(\omega+i\eta) \nonumber \\
& = & \lambda^{-1+\alpha} A_\alpha(\omega).
\end{eqnarray}
Consequently, we have
\begin{eqnarray}
\bar{\kappa}_{\alpha}\left(\lambda\epsilon,\lambda\epsilon^{\prime}\right) & = & \lambda^{-1+2\alpha}\bar{\kappa}_{\alpha}\left(\epsilon,\epsilon^{\prime}\right), \\
\tilde{S}_{\alpha}^{\pr\pr}\left(\lambda\epsilon_{1}\lambda\epsilon_{2},\lambda\epsilon_{3},\lambda\omega\right) & = & \lambda^{-1+2\alpha}\tilde{S}_{\alpha}^{\pr\pr}\left(\epsilon_{1},\epsilon_{2},\epsilon_{3},\omega\right).
\end{eqnarray}
Then, approximating the density of states to be constant near the Fermi level,
we find that the imaginary part of the electron self-energy in the $T\rightarrow0$
limit becomes

\begin{eqnarray}
\Sigma^{\prime\prime}\left(k,\omega\right) & = & -U^2\sum_{p_{1}p_{2}p_{3}}\delta_{p_{1}+p_{3},p_{2}+k}\tilde{S}_{\alpha}^{\pr\pr}\left(\epsilon_{p_{1}},\epsilon_{p_{2}},E_{p_{3}},\omega\right)\nonumber \\
 & \sim & -U^2 \int d\epsilon_{1}d\epsilon_{2}dE\ \tilde{S}_{\alpha}^{\pr\pr}\left(\epsilon_{1},\epsilon_{2},E,\omega\right),
\end{eqnarray}
which scales with respect to energy $\omega$ as 
\begin{eqnarray}
\Sigma^{\prime\prime}\left(k,\lambda\omega\right) & = & \lambda^{2+2\alpha}\Sigma^{\prime\prime}\left(k,\omega\right).
\end{eqnarray}
Therefore, the electron self-energy due to electron-unparticle interactions
behaves as $\Sigma^{\prime\prime}\sim\omega^{2+2\alpha}$ at low temperatures,
deviating from the Fermi liquid behavior of $\Sigma_{\text{FL}}^{\prime\prime}\sim\omega^{2}$.
In the $\omega\rightarrow0$ limit, a similar argument shows that
$\Sigma^{\prime\prime}\sim T^{2+2\alpha}$ at low energies. Summarized
in Fig. \ref{fig:self-energy-plot}, these scaling behaviors of the
electron self-energy are our main result; they hold for $-1<\alpha<1$,
and do not depend on the specific form of the electron
energy spectrum, $E_{k}$. For $\alpha\lesssim0$,
this non-Fermi-liquid state of matter quantitatively corresponds to
the power-law liquid revealed in the cuprates by the recent ARPES
measurements \cite{Reber2015}.

\subsection{Susceptibility}

\begin{figure}[H]
\begin{centering}
\includegraphics[width=0.8\columnwidth]{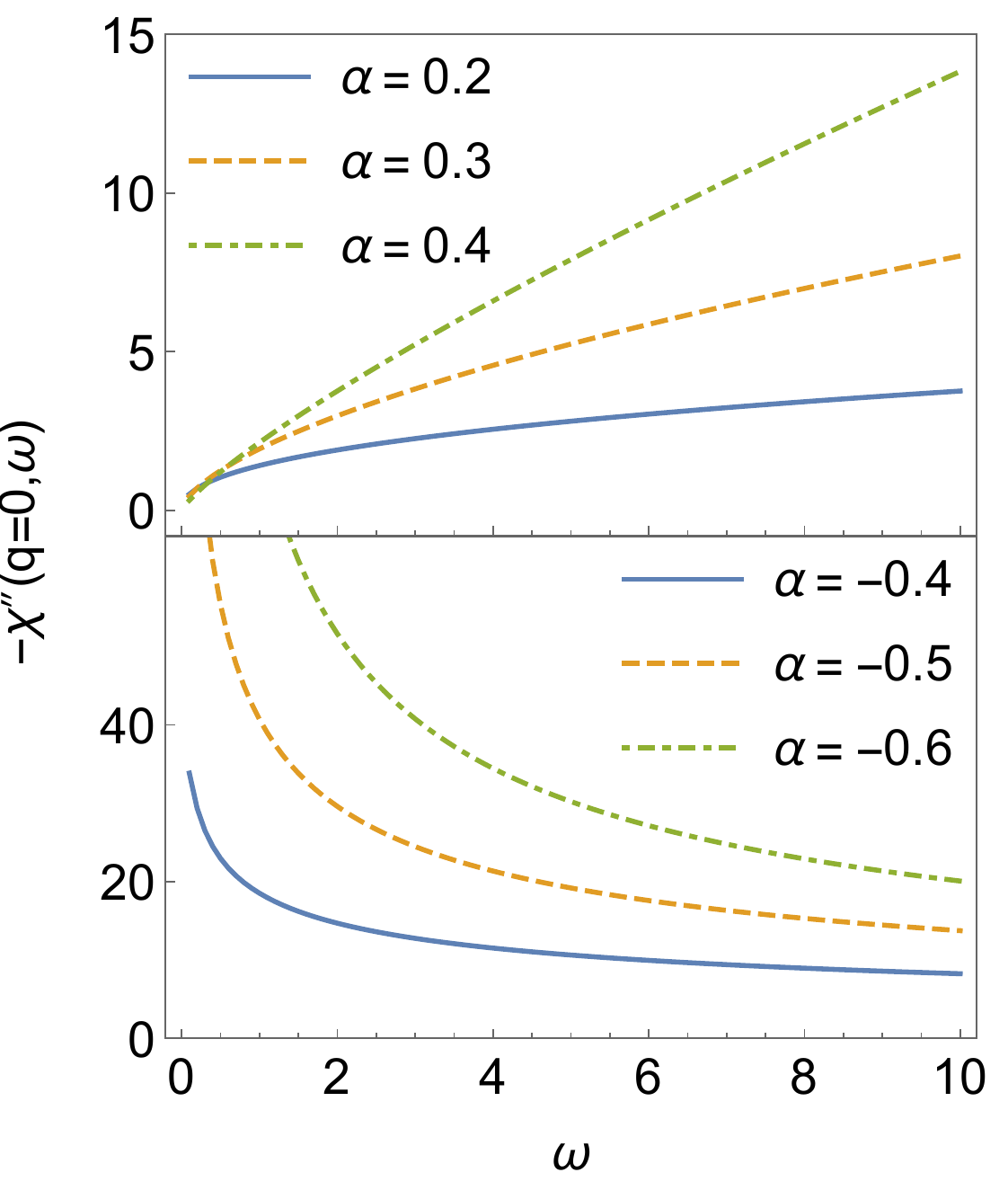}
\par\end{centering}
\caption{The energy dependence of the unparticle susceptibility $\chi_{\alpha}^{\protect\pr\protect\pr}$
for a quadratic energy spectrum $\epsilon_{k}\sim p^{2}$ in the $T\rightarrow0$
and $q\rightarrow0$ limit, for various values of $\alpha$. The scaling
behavior $\chi_{\alpha}^{\protect\pr\protect\pr}\sim\omega^{2\alpha}$
is associated with the scaling of the electron self-energy depicted
in Fig. \ref{fig:self-energy-plot}. Note that these plots are only
qualitatively accurate due to issues with numerical stability. \label{fig:susceptibility-vs-energy}}
\end{figure}

The scaling behavior of the electron self-energy can be traced back
to the unparticle susceptibility $\chi_{\alpha}$ defined by Eq. \ref{eq:suscep}.
From the calculations in Appendix \ref{app:Matsubara-sums}, its imaginary
part after analytic continuation $i\omega_{n}\rightarrow\omega+i\eta$
can be written as 
\begin{eqnarray}
\chi_{\alpha}^{\pr\pr}\left(q,\omega\right) & = & \sum_{p}S_{\alpha}^{\pr\pr}\left(\epsilon_{p},\epsilon_{p-q},\omega\right),
\end{eqnarray}
where

\begin{eqnarray}
S_{\alpha}^{\pr\pr}\left(\epsilon_{1},\epsilon_{2},\omega\right) & = & \bar{\kappa}_{\alpha}\left(\epsilon_{1},\epsilon_{2}+\omega\right)-\bar{\kappa}_{\alpha}\left(\epsilon_{2},\epsilon_{1}-\omega\right).
\end{eqnarray}
The function $\bar{\kappa}_{\alpha}$ is defined above by Eq. \ref{eq:kappa-def}.
Fig. \ref{fig:susceptibility-vs-energy} illustrates the unparticle
susceptibility in the $q\rightarrow0$ and $T\rightarrow0$ limit
for a quadratic energy spectrum $\epsilon_{k}\sim k^{2}$ in two dimensions.
We note three features distinctive from the analogous free electron
susceptibility. First, the unparticle susceptibility is nonzero despite
the chemical potential being restricted to be zero on account of scale invariance. This is unlike normal particles for which a zero chemical potential necessarily implies that there is zero filling and hence zero susceptibility.
Second, the unparticle susceptibility does not have a cutoff at high energies. Third, from 
\begin{eqnarray}
\chi^{\pr\pr}\left(q=0,\lambda\omega\right) & = & \int d^{2}pS_{\alpha}^{\pr\pr}\left(p^{2},p^{2},\lambda\omega\right)\nonumber \\
 & = & \pi\int_{0}^{\infty}d\epsilon S_{\alpha}^{\pr\pr}\left(\epsilon,\epsilon,\lambda\omega\right).\nonumber \\
 & = & \lambda^{2\alpha}\chi_{\alpha}^{\pr\pr}\left(0,\lambda\omega\right),
\end{eqnarray}
we see that the susceptibility scales as $\chi^{\pr\pr}\left(0,\omega\right)\sim\omega^{2\alpha}$. Such a scaling form ensures that when $\alpha<0$ ($\alpha>0$), the susceptibility is enhanced (suppressed) at low energies, as shown in Fig. \ref{fig:susceptibility-vs-energy}. Such an enhancement (suppression) is crucial for the increased (decreased) scattering rate, as quantified by the electron self-energy in the previous subsection. These features completely violate the usual susceptibility sum rule
and can be attributed to the broadening of the unparticle spectral function. As $\left|\alpha\right|$ decreases, the features become less pronounced, as expected.

Similar non-Fermi liquid behavior induced by the enhancement of low energy susceptibility also occurs, for example, in systems where large portions of the Fermi surface are nested with a single nesting wave vector \cite{Virosztek1990, Virosztek1999}, and in  multiband models with orbital fluctuations \cite{Lee2012}. Additionally, the self-energy of a Fermi liquid in the presence of weak impurities has an imaginary part of the form $\Sigma''\sim(E-E_f)^{d/2}$, where $d$ is the spatial dimension\cite{Giuliani2005}.  Non-Fermi liquid behavior in this case can also be understood as an enhancement in the low energy spectrum of the susceptibility\cite{Giuliani2005}.

\subsection{Occupancy}

\begin{figure}[H]
\begin{centering}
\includegraphics[width=1.0\columnwidth]{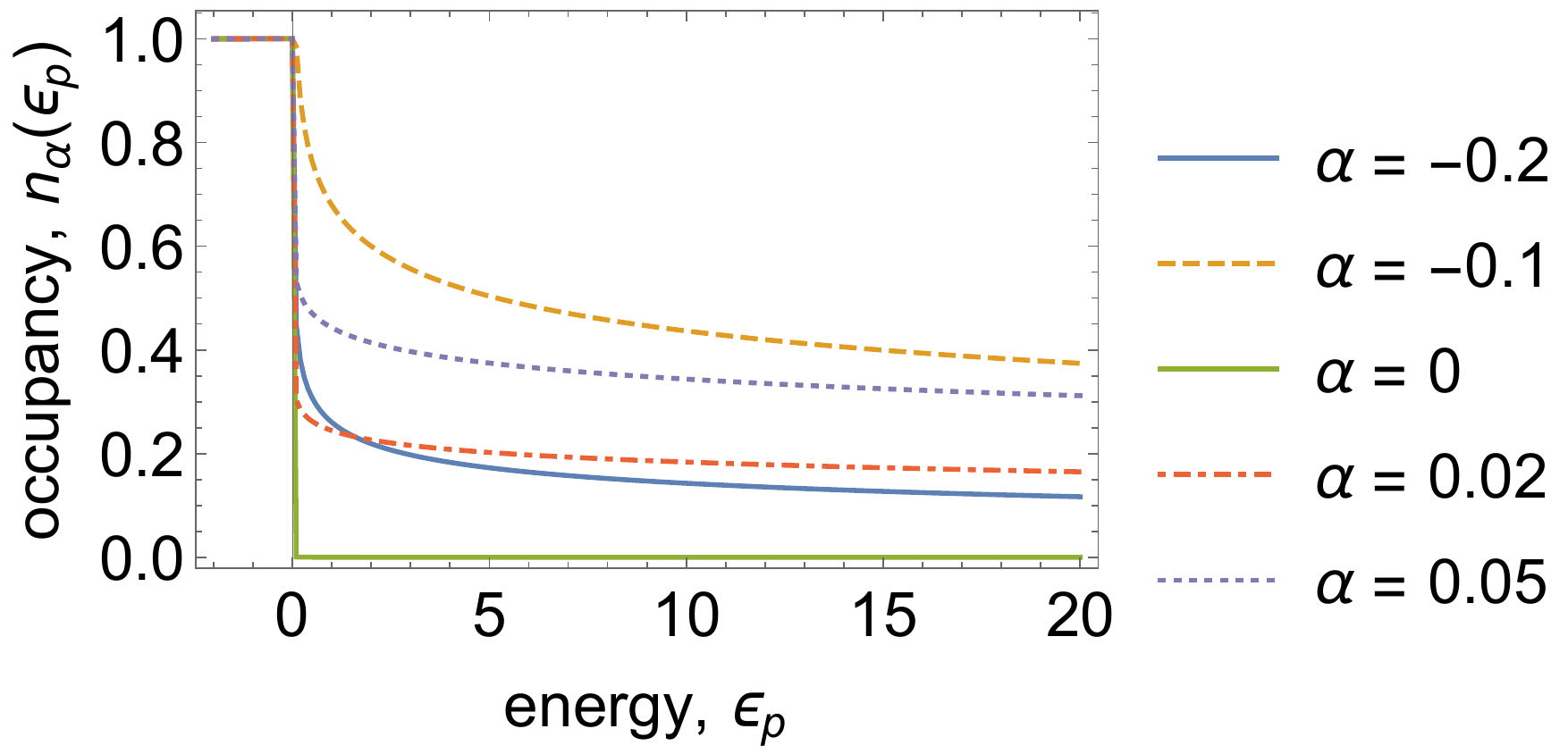}
\par\end{centering}
\caption{The energy $\epsilon_{p}$ dependence of the unparticle occupancy
at $T=0$. The significant occupancy at large $\epsilon_{p}$ differs
from the Fermi distribution.\label{fig:occupancy-vs-energy}}
\end{figure}

In Fermi liquid theory, the quadratic scaling of the electron self-energy
follows from a phase-space argument involving the occupancy
of electrons. Therefore, it can be illuminating to explore how this
argument is modified in the case of unparticles by computing the
occupancy for unparticles, 
\begin{eqnarray}
n_{\alpha}\left(\epsilon_{p}\right) & = & \int_{-\infty}^{\infty}dz\ n_{F}\left(z\right)A_{\alpha}\left(z-\epsilon_{p}\right)e^{z0^{+}}.
\end{eqnarray}
Fig. \ref{fig:occupancy-vs-energy} shows that in the $T\rightarrow0$
limit, unlike the Fermi distribution for particles, the occupancy
of unparticles is significant even when $\epsilon_{p}$ is large.
This counterintuitive result can be understood by noting that the
occupancy number measures the filling of states at momentum $p$,
instead of at energy $\epsilon_{p}$. This distinction is important
because, unlike the particle case, the unparticle spectral function is
broadened over a wide energy range. Consequently, even unparticles
with a large $\epsilon_{p}$ possess a significant amount of low energy
states that are filled at low temperatures. For $\alpha<0$, these states enlarge the
scattering phase space in the electron self-energy by enhancing the low energy susceptibility bubble, resulting in the
non-Fermi liquid behavior described in the preceding section.   In addition,
the occupancy is notably non-symmetric, reflecting the particle-hole
asymmetry of the unparticle Green function. This enhancement of phase space undoubtedly reflects the enhanced scattering rate that ultimately grows linearly with temperature as opposed to the standard $T^2$ in the Fermi liquid case.

\section{Discussion and conclusions}

As discussed in Ref. \onlinecite{Reber2015}, a sublinear scaling
of the electron self-energy can be interpreted as having a vanishing
quasiparticle residue $Z$ in Fermi liquid theory. This signifies
that interactions with fermionic unparticles with $\alpha<-0.5$ cause
electrons to behave completely incoherently, which is unsurprising
given the nature of unparticles. Nevertheless, since $\Sigma^{\prime\prime}\left(\omega=0,T=0\right)=0$,
the Fermi surface remains sharp \cite{Senthil2008}.

We can similarly calculate the self-energy of unparticles due to self
interactions, that is, when the electron line in Fig. \ref{fig:self-energy_feynman-diagram}
is replaced by another unparticle line. Naively, we expect the self-energy
to scale as $\Sigma^{\prime\prime}\sim\omega^{2+3\alpha}$ and $T^{2+3\alpha}$.
This result, as well as the susceptibility and occupancy calculated
above, can in principle be observed experimentally. However, any meaningful
comparison with experimental observations would require further knowledge
about the form of couplings between unparticles and external fields.

Our recent paper \cite{Limtragool2016} studied the effects of bosonic
unparticles on the electron self-energy. While similar scaling behaviors
were obtained, there are a few subtle differences. First, while a unitarity bound constrains the scaling dimension of bosonic unparticles, we do not know of any such constraint for fermionic unparticles. This freedom allows for a more qualitative agreement with experiments.
 Second, unlike the results in the bosonic case, there is no dependence on the dimensionality in the scaling of $\Sigma''$. This state of affairs obtains because we approximate the density of states of both electrons and fermionic unparticles to be constant near the Fermi level. Third, our susceptibility plots in Fig. \ref{fig:susceptibility-vs-energy} differ from that in Ref. \onlinecite{Limtragool2016}, because a nonzero chemical potential was previously adopted.

In conclusion, we showed analytically that interactions between electrons
and fermionic unparticles---a scale-invariant state of matter---can produce
the power-law liquid revealed in the cuprates by recent ARPES experiments
\cite{Reber2015}. In particular, we found that, at low temperatures
and energies, the electron self-energy due to interactions with fermionic
unparticles exhibits power-law scaling with respect to energy and
temperature: $\Sigma^{\prime\prime}\sim\omega^{2+2\alpha}$ and $T^{2+2\alpha}$,
where $\alpha$ is the anomalous scaling of the unparticle propagator.
This non-Fermi-liquid behavior can be attributed to the broadening
of the unparticle spectral function over a wide energy range, which
drastically alters the scattering phase space by enhancing (suppressing) the intrinsic susceptibility at low energies for negative (positive) $\alpha$. Although unparticles
have zero chemical potential as required by scale invariance, they
nevertheless can contribute to the electron self-energy due to the
same broadening. Our results present new evidence suggesting that
unparticles might be important low-energy degrees of freedom in the
cuprates, and should inspire the interpretation of other experimental data
using unparticles.
\begin{acknowledgments}
We thank the NSF DMR-1461952 for partial funding of this project.
ZL is supported by the Department of Physics at the University
of Illinois and a scholarship from the Agency of Science, Technology
and Research. CS and PWP are supported by the Center for Emergent
Superconductivity, a DOE Energy Frontier Research Center, Grant No.
DE-AC0298CH1088. KL is supported by the Department of Physics at the
University of Illinois and a scholarship from the Ministry of Science
and Technology, Royal Thai Government. 
\end{acknowledgments}

\onecolumngrid

\appendix

\section{Analytic evaluation of Matsubara sums \label{app:Matsubara-sums}}

\subsection{Susceptibility}

\begin{figure}[H]
\begin{centering}
\subfloat[\label{fig:matsum-contour}]{\begin{centering}
\includegraphics[width=0.3\columnwidth]{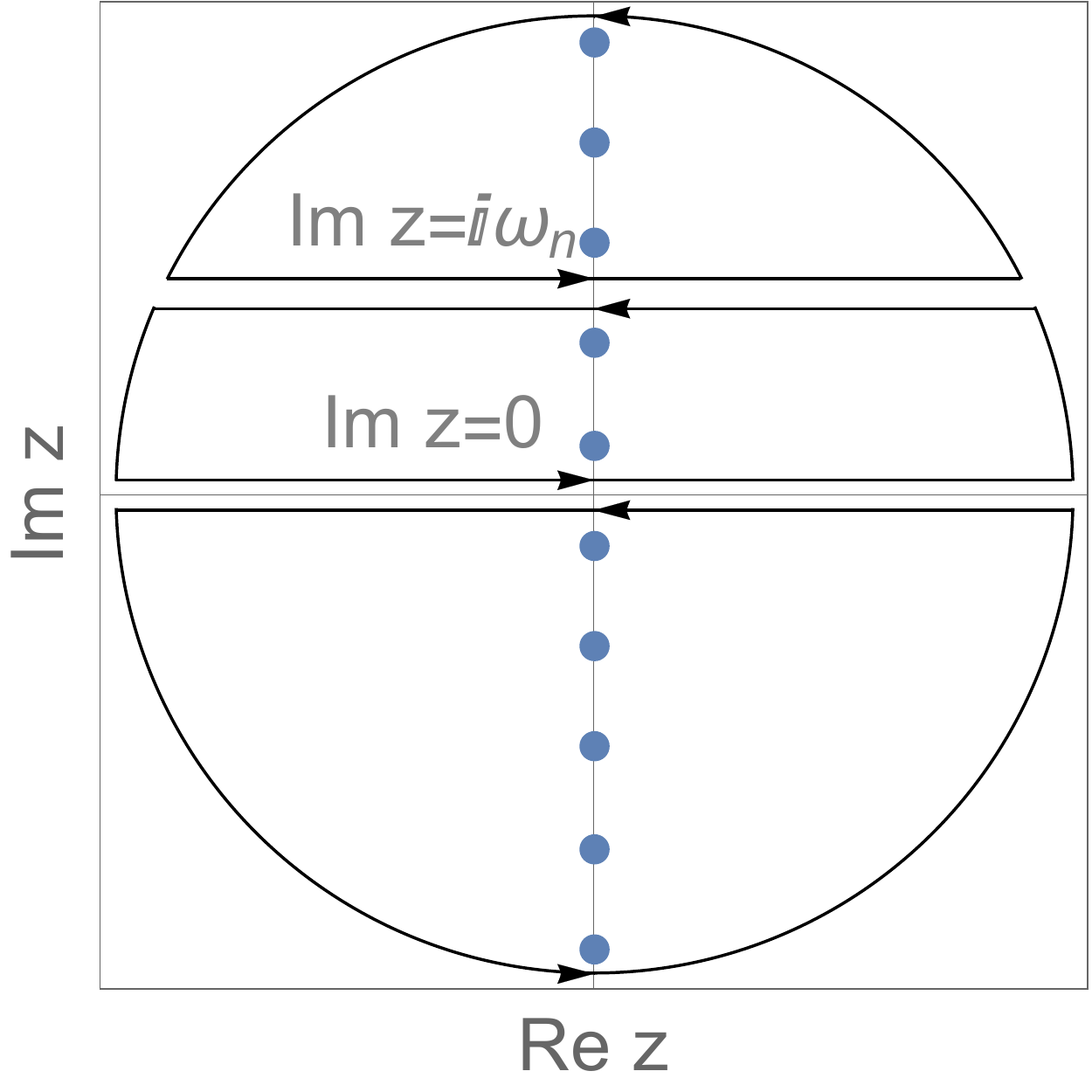}
\par\end{centering}

}\qquad{}\subfloat[\label{fig:matsum-contour-2}]{\begin{centering}
\includegraphics[width=0.3\columnwidth]{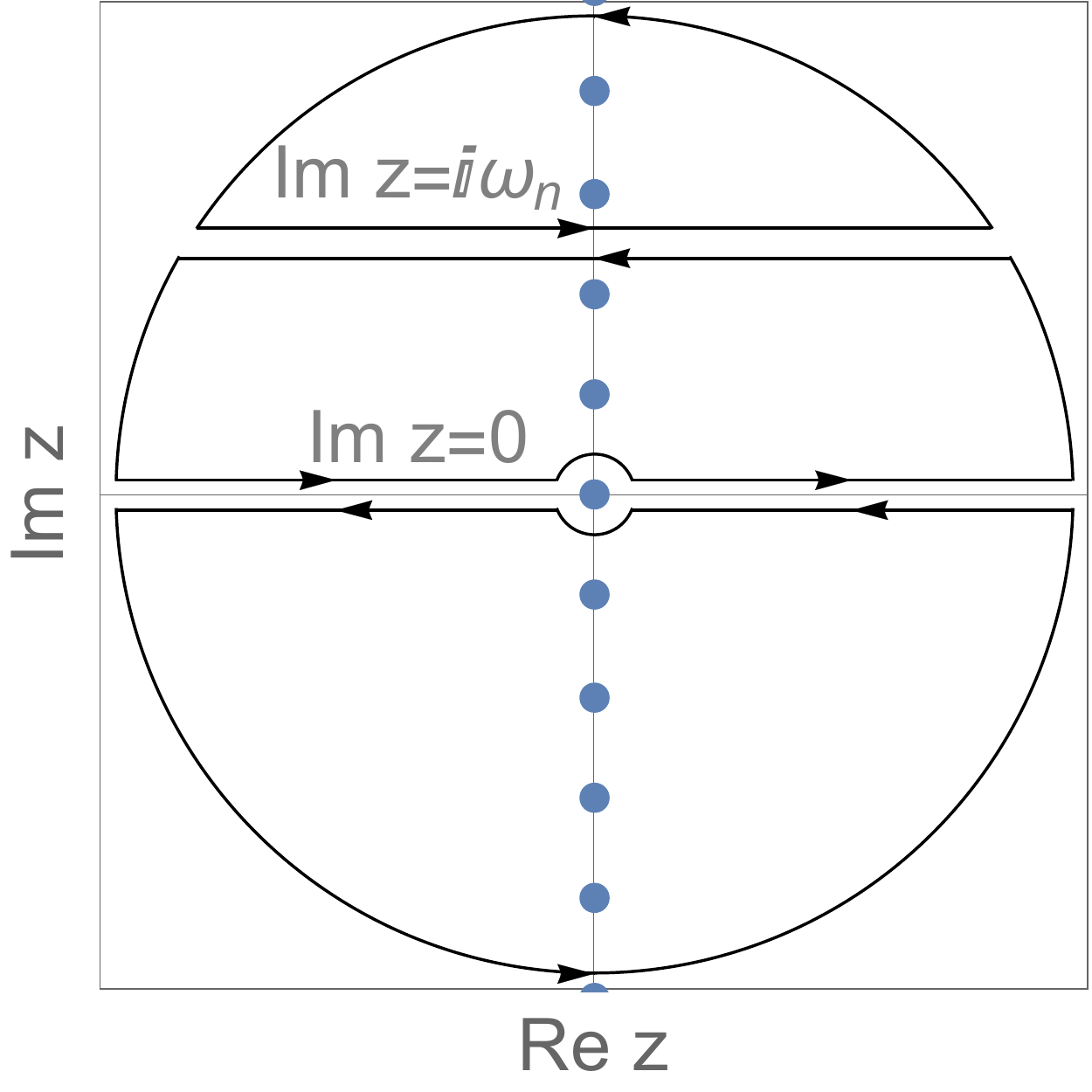}
\par\end{centering}
}
\par\end{centering}
\caption{The contours used to evaluate the Matsubara sums in (a) the unparticle
susceptibility, and (b) the electron self-energy.}
\end{figure}

The unparticle susceptibility defined by Eq. \ref{eq:suscep} involves
the fermionic Matsubara sum

\begin{eqnarray*}
S_\alpha\left(\epsilon_{1},\epsilon_{2},i\omega_{n}\right) & = & T\sum_{i\omega_{m}}G_{\alpha}\left(i\omega_{m}-\epsilon_{1}\right)G_{\alpha}\left(i\omega_{m}-i\omega_{n}-\epsilon_{2}\right),
\end{eqnarray*}
where $i\omega_{n}$ is a bosonic Matsubara frequency. Using Cauchy's
residue theorem, we rewrite the Matsubara sum as 
\begin{eqnarray*}
S_\alpha\left(\epsilon_{1},\epsilon_{2},i\omega_{n}\right) & = & -\frac{1}{2\pi i}\oint_{C}dz\ n_{F}\left(z\right)G_{\alpha}\left(z-\epsilon_{1}\right)G_{\alpha}\left(z-i\omega_{n}-\epsilon_{2}\right),
\end{eqnarray*}
where $n_{F}\left(z\right)=\left(e^{z/T}+1\right)^{-1}$ is the Fermi
distribution. Since the integrand is analytic except along $\im z=0$
and $\im z=i\omega_{n}$, we use the contour $C$ illustrated in Fig.
\ref{fig:matsum-contour}.

The integrals along the paths at large radius vanish when $\alpha<1/2$. For $\alpha\geq 1/2$, a convergence factor $e^{z0^+}$ can be included so that the same integrals vanish. Consequently, the nonvanishing contributions to the contour integral are those along
the branch cuts: 

\begin{eqnarray*}
I_{b1}\left(\epsilon_{1},\epsilon_{2},i\omega_{n}\right) & = & -\frac{1}{2\pi i}\int_{-\infty}^{\infty}dz\ n_{F}\left(z\right)\left[G_{\alpha}\left(z^{+}-\epsilon_{1}\right)-G_{\alpha}\left(z^{-}-\epsilon_{1}\right)\right]G_{\alpha}\left(z-i\omega_{n}-\epsilon_{2}\right)\\
 & = & -\frac{1}{2\pi i}\int_{-\infty}^{\infty}dz\ n_{F}\left(z\right)2i\im\left[G_{\alpha}\left(z^{+}-\epsilon_{1}\right)\right]G_{\alpha}\left(z-i\omega_{n}-\epsilon_{2}\right)\\
 & = & \int_{-\infty}^{\infty}dzn_{F}\left(z\right)A_{\alpha}\left(z-\epsilon_{1}\right)G_{\alpha}\left(z-i\omega_{n}-\epsilon_{2}\right),\\
I_{b2}\left(\epsilon_{1},\epsilon_{2},i\omega_{n}\right) & = & -\frac{1}{2\pi i}\int_{-\infty+i\omega_{n}}^{\infty+i\omega_{n}}dz\ n_{F}\left(z\right)G_{\alpha}\left(z-\epsilon_{1}\right)\left[G_{\alpha}\left(z^{+}-i\omega_{n}-\epsilon_{2}\right)-G_{\alpha}\left(z^{-}-i\omega_{n}-\epsilon_{2}\right)\right]\\
 & = & -\frac{1}{2\pi i}\int_{-\infty}^{\infty}dz\ n_{F}\left(z+i\omega_{n}\right)G_{\alpha}\left(z+i\omega_{n}-\epsilon_{1}\right)\left[G_{\alpha}\left(z^{+}-\epsilon_{2}\right)-G_{\alpha}\left(z^{-}-\epsilon_{2}\right)\right]\\
 & = & -\frac{1}{2\pi i}\int_{-\infty}^{\infty}dz\ n_{F}\left(z\right)G_{\alpha}\left(z+i\omega_{n}-\epsilon_{1}\right)2i\im G_{\alpha}\left(z^{+}-\epsilon_{2}\right)\\
 & = & \int_{-\infty}^{\infty}dz\ n_{F}\left(z\right)G_{\alpha}\left(z+i\omega_{n}-\epsilon_{1}\right)A_{\alpha}\left(z-\epsilon_{2}\right).
\end{eqnarray*}
Here $z^{\pm} = z \pm i\eta$, with $\eta = 0^+$ being a positive real infinitesimal. After analytic continuation $i\omega_{n}\rightarrow\omega+i\eta$,
the imaginary part of the Matsubara sum becomes

\begin{eqnarray*}
\im S_{\alpha}\left(\epsilon_{1},\epsilon_{2},\omega+i\eta\right) & = & \int_{-\infty}^{\infty}dz\ n_{F}\left(z\right)\left[A_{\alpha}\left(z-\epsilon_{1}\right)\im G_{\alpha}\left(z-\omega-i\eta-\epsilon_{2}\right)+\im G_{\alpha}\left(z+\omega+i\eta-\epsilon_{1}\right)A_{\alpha}\left(z-\epsilon_{2}\right)\right]\\
 & = & \int_{-\infty}^{\infty}dz\ n_{F}\left(z\right)\left[A_{\alpha}\left(z-\epsilon_{1}\right)\pi A_{\alpha}\left(z-\omega-\epsilon_{2}\right)-\pi A_{\alpha}\left(z+\omega-\epsilon_{1}\right)A_{\alpha}\left(z-\epsilon_{2}\right)\right]\\
 & = & \pi\int_{-\infty}^{\infty}dz\ n_{F}\left(z\right)A_{\alpha}\left(z-\epsilon_{1}\right)A_{\alpha}\left(z-\omega-\epsilon_{2}\right)-\left(\epsilon_{1}\leftrightarrow\epsilon_{2},\omega\rightarrow-\omega\right)\\
 & \equiv & \bar{\kappa}_{\alpha}\left(\epsilon_{1},\epsilon_{2}+\omega\right)-\bar{\kappa}_{\alpha}\left(\epsilon_{2},\epsilon_{1}-\omega\right),
\end{eqnarray*}
where we have defined
\begin{eqnarray*}
\bar{\kappa}_{\alpha}\left(\epsilon,\epsilon^{\pr}\right) & = & \pi\int_{-\infty}^{\infty}dz\ n_{F}\left(z\right)A_{\alpha}\left(z-\epsilon\right)A_{\alpha}\left(z-\epsilon^{\pr}\right).
\end{eqnarray*}
For $\alpha>0$, we can evaluate this exactly in the $T\rightarrow0$
limit using the unparticle spectral function in Eq. \ref{eq:spectral_func}:
\begin{eqnarray*}
\bar{\kappa}_{\alpha}\left(\epsilon,\epsilon^{\prime}\right) & = & \frac{1}{\pi}\sin^{2}\left(\pi\alpha\right)\int_{-\infty}^{\min\left(\epsilon,0,\epsilon^{\prime}\right)}dz\frac{1}{\left(\epsilon-z\right)^{1-\alpha}}\frac{1}{\left(\epsilon^{\prime}-z\right)^{1-\alpha}}\\
 & = & \frac{1}{\pi}\sin^{2}\left(\pi\alpha\right)2^{1-2\alpha}\int_{\xi\left(\epsilon,\epsilon^{\prime}\right)}^{\infty}\frac{dz}{\left[z^{2}-\left(\epsilon-\epsilon^{\prime}\right)^{2}\right]^{1-\alpha}}\\
 & = & \frac{1}{\pi}\sin^{2}\left(\pi\alpha\right)\frac{1}{2}\left[\frac{2}{\xi\left(\epsilon,\epsilon^{\prime}\right)}\right]^{1-2\alpha}\int_{0}^{1}dt\frac{t^{-\frac{1}{2}-\alpha}}{\left[1-t\left|\frac{\epsilon-\epsilon^{\prime}}{\xi\left(\epsilon,\epsilon^{\prime}\right)}\right|^{2}\right]^{1-\alpha}}\\
 & = & \frac{1}{\pi}\sin^{2}\left(\pi\alpha\right)\frac{1}{1-2\alpha}\left[\frac{2}{\xi\left(\epsilon,\epsilon^{\prime}\right)}\right]^{1-2\alpha}\hf\left[1-\alpha,\frac{1}{2}-\alpha;\frac{3}{2}-\alpha;\left|\frac{\epsilon-\epsilon^{\prime}}{\xi\left(\epsilon,\epsilon^{\prime}\right)}\right|^{2}\right],
\end{eqnarray*}
where $\xi\left(\epsilon,\epsilon^{\prime}\right)=\max\left(\left|\epsilon-\epsilon^{\prime}\right|,\epsilon+\epsilon^{\prime}\right)$,
and $\hf\left(a,b;c,z\right)$ is the hypergeometric function.

\subsection{Self-energy}

The electron self-energy defined by Eq. \ref{eq:self-energy} involves
the bosonic Matsubara sum 

\begin{eqnarray*}
\tilde{S}_{\alpha}\left(\epsilon_{1},\epsilon_{2},\epsilon_{3},i\omega_{n}\right) & = & T\sum_{i\omega_{m^{\prime}}}G_{0}\left(i\omega_{n}-i\omega_{m^{\pr}}-\epsilon_{3}\right)S_{\alpha}\left(\epsilon_{1},\epsilon_{2},i\omega_{m^{\pr}}\right)\\
 & = & \frac{1}{2\pi i}\oint_{C^{\pr}}dz^{\pr}n_{B}\left(z^{\pr}\right)G_{0}\left(i\omega_{n}-z^{\pr}-\epsilon_{3}\right)S_{\alpha}\left(\epsilon_{1},\epsilon_{2},z^{\pr}\right) + T G_{0}\left(i\omega_{n}-\epsilon_{3}\right)S_{\alpha}\left(\epsilon_{1},\epsilon_{2},0\right),
\end{eqnarray*}
where $i\omega_{n}$ is a fermionic Matsubara frequency, and $n_{B}\left(z\right)=\left(e^{z/T}-1\right)^{-1}$
is the Bose distribution. Since the integrand has a branch cut along
$\im z^{\pr}=0$ and a pole on a line $\im z^{\pr}=i\omega_{n}$, we adopt the contour
$C^{\prime}$ shown in Fig. \ref{fig:matsum-contour-2}. If $\alpha<1$, the integrals along the paths at large radius vanish.

The integrals along the small circle of radius $r$ around the origin require special consideration. Since $S_{\alpha}\left(\epsilon_{1},\epsilon_{2},z^{\pr}\right)$  is analytic in the upper (lower) half plane, we see that $S_{\alpha}\left(\epsilon_{1},\epsilon_{2},z^{\pr}\right) \rightarrow S_{\alpha}\left(\epsilon_{1},\epsilon_{2},0\right)$ in the limit $z^{\pr}\rightarrow 0$ for $z^\pr$ in the same domain. Hence, as the radius $r\rightarrow 0$, the integral along the small circle reduces to
\begin{eqnarray*}
\frac{1}{2\pi i}\oint_{|z^\pr|=r}dz^{\pr}n_{B}\left(z^{\pr}\right)G_{0}\left(i\omega_{n}-z^{\pr}-\epsilon_{3}\right)S_{\alpha}\left(\epsilon_{1},\epsilon_{2},z^{\pr}\right) & \sim & \left[\frac{1}{2\pi i}\oint_{|z^\pr|=r}dz^{\pr}n_{B}\left(z^{\pr}\right)\right] G_{0}\left(i\omega_{n}-\epsilon_{3}\right)S_{\alpha}\left(\epsilon_{1},\epsilon_{2},0\right) \\
& = & - T G_{0}\left(i\omega_{n}-\epsilon_{3}\right)S_{\alpha}\left(\epsilon_{1},\epsilon_{2},0\right),
\end{eqnarray*}
which exactly cancels the $i\omega_{m^{\pr}}=0$ term in $\tilde{S}_{\alpha}\left(\epsilon_{1},\epsilon_{2},\epsilon_{3},i\omega_{n}\right)$. This cancellation can be physically motivated. First, notice that the imaginary part of the term contains the factor $\delta(\omega-\epsilon_3)$ after analytic continuation. Then, since $\omega=\epsilon_3$ corresponds to no energy transfer between unparticles and electrons, such a term understandably should not contribute to the electron self-energy.

Then, the nonvanishing contributions to $\tilde{S}_\alpha$ are simply the integrals along the lines $\im z^{\pr}=0$ and $\im z^{\pr}=i\omega_{n}$:

\begin{eqnarray*}
\tilde{I}_{b1}\left(\epsilon_{1},\epsilon_{2},\epsilon_{3},i\omega_{n}\right) & = & \frac{1}{2\pi i}\int_{-\infty}^{\infty}dz^{\pr}n_{B}\left(z^{\pr}\right)G_{0}\left(i\omega_{n}-z^{\pr},\epsilon_{3}\right)\left[S_{\alpha}\left(\epsilon_{1},\epsilon_{2},z^{\pr+}\right)-S_{\alpha}\left(\epsilon_{1},\epsilon_{2},z^{\pr-}\right)\right]\\
 & = & \frac{1}{2\pi i}\int_{-\infty}^{\infty}dz^{\pr}n_{B}\left(z^{\pr}\right)G_{0}\left(i\omega_{n}-z^{\pr},\epsilon_{3}\right)2i\im S_{\alpha}\left(\epsilon_{1},\epsilon_{2},z^{\pr+}\right)\\
 & = & \frac{1}{\pi}\int_{-\infty}^{\infty}dz^{\pr}n_{B}\left(z^{\pr}\right)G_{0}\left(i\omega_{n}-z^{\pr},\epsilon_{3}\right)\left[\bar{\kappa}_{\alpha}\left(\epsilon_{1},\epsilon_{2}+z^{\pr}\right)-\bar{\kappa}_{\alpha}\left(\epsilon_{2},\epsilon_{1}-z^{\pr}\right)\right],\\
\tilde{I}_{b2}\left(\epsilon_{1},\epsilon_{2},\epsilon_{3},i\omega_{n}\right) & = & \frac{1}{2\pi i}\int_{-\infty+i\omega_{n}}^{\infty+i\omega_{n}}dz^{\pr}n_{B}\left(z^{\pr}\right)\left[G_{0}\left(i\omega_{n}-z^{\pr+},\epsilon_{3}\right)-G_{0}\left(i\omega_{n}-z^{\pr-},\epsilon_{3}\right)\right]S_{\alpha}\left(\epsilon_{1},\epsilon_{2},z^{\pr}\right)\\
 & = & \frac{1}{2\pi i}\int_{-\infty}^{\infty}dz^{\pr}n_{B}\left(z^{\pr}+i\omega_{n}\right)\left[G_{0}\left(-z^{\pr+},\epsilon_{3}\right)-G_{0}\left(-z^{\pr-},\epsilon_{3}\right)\right]S_{\alpha}\left(\epsilon_{1},\epsilon_{2},z^{\pr}+i\omega_{n}\right)\\
 & = & \frac{1}{2\pi i}\int_{-\infty}^{\infty}dz^{\pr}\left[-n_{F}\left(z^{\pr}\right)\right]2i\im\left[G_{0}\left(-z^{\pr+},\epsilon_{3}\right)\right]S_{\alpha}\left(\epsilon_{1},\epsilon_{2},z^{\pr}+i\omega_{n}\right)\\
 & = & -\frac{1}{2\pi i}\int_{-\infty}^{\infty}dz^{\pr}n_{F}\left(z^{\pr}\right)2i\pi A_{0}\left(-z^{\pr},\epsilon_{3}\right)S_{\alpha}\left(\epsilon_{1},\epsilon_{2},z^{\pr}+i\omega_{n}\right)\\
 & = & -\int_{-\infty}^{\infty}dz^{\pr}n_{F}\left(z^{\pr}\right)A_{0}\left(-z^{\pr},\epsilon_{3}\right)S_{\alpha}\left(\epsilon_{1},\epsilon_{2},z^{\pr}+i\omega_{n}\right).
\end{eqnarray*}
Here, $A_0 = -\frac{1}{\pi}\mathrm{Im}G_0$ is the spectral function of $G_0$ , and we take principal values of the integrals in $\tilde{I}_{b1}$ due to the cancellation mentioned above. Then, analytic continuation
$i\omega_{n}\rightarrow\omega+i\eta$ gives
\begin{eqnarray*}
\im\tilde{I}_{b1}\left(\epsilon_{1},\epsilon_{2},\epsilon_{3},\omega+i\eta\right) & = & -\int_{-\infty}^{\infty}dz^{\pr}n_{B}\left(z^{\pr}\right)A_{0}\left(\omega-z^{\pr},\epsilon_{3}\right)\left[\bar{\kappa}_{\alpha}\left(\epsilon_{1},\epsilon_{2}+z^{\prime}\right)-\bar{\kappa}_{\alpha}\left(\epsilon_{2},\epsilon_{1}-z^{\prime}\right)\right],\\
\im\tilde{I}_{b2}\left(\epsilon_{1},\epsilon_{2},\epsilon_{3},\omega+i\eta\right) & = & -\int_{-\infty}^{\infty}dz^{\pr}n_{F}\left(z^{\pr}\right)A_{0}\left(-z^{\pr},\epsilon_{3}\right)\left[\bar{\kappa}_{\alpha}\left(\epsilon_{1},\epsilon_{2}+z^{\pr}+\omega\right)-\bar{\kappa}_{\alpha}\left(\epsilon_{2},\epsilon_{1}-z^{\pr}-\omega\right)\right]\\
 & = & -\int_{-\infty}^{\infty}dz^{\pr}n_{F}\left(z^{\pr}-\omega\right)A_{0}\left(\omega-z^{\pr},\epsilon_{3}\right)\left[\bar{\kappa}_{\alpha}\left(\epsilon_{1},\epsilon_{2}+z^{\pr}\right)-\bar{\kappa}_{\alpha}\left(\epsilon_{2},\epsilon_{1}-z^{\pr}\right)\right],\\
\im\tilde{S}\left(\epsilon_{1},\epsilon_{2},\epsilon_{3},\omega+i\eta\right) & = & -\int_{-\infty}^{\infty}dz^{\pr}\left[n_{B}\left(z^{\pr}\right)+n_{F}\left(z^{\pr}-\omega\right)\right]A_{0}\left(\omega-z^{\pr},\epsilon_{3}\right)\left[\bar{\kappa}_{\alpha}\left(\epsilon_{1},\epsilon_{2}+z^{\prime}\right)-\bar{\kappa}_{\alpha}\left(\epsilon_{2},\epsilon_{1}-z^{\prime}\right)\right]\\
 & = & \int_{-\infty}^{\infty}dz^{\pr}A_{0}\left(\omega-z^{\pr},\epsilon_{3}\right)\left[n_{B}\left(z^{\pr}\right)+n_{F}\left(z^{\pr}-\omega\right)\right]\left[\bar{\kappa}_{\alpha}\left(\epsilon_{2},\epsilon_{1}-z^{\prime}\right)-\bar{\kappa}_{\alpha}\left(\epsilon_{1},\epsilon_{2}+z^{\prime}\right)\right].
\end{eqnarray*}
Finally, using $A_{0}\left(\omega\right)=\delta\left(\omega\right)$
gives 
\begin{eqnarray*}
\im\tilde{S}_{\alpha}\left(\epsilon_{1},\epsilon_{2},\epsilon_{3},\omega+i\eta\right) & = & \left[n_{B}\left(\omega-\epsilon_{3}\right)+n_{F}\left(-\epsilon_{3}\right)\right]\left[\bar{\kappa}_{\alpha}\left(\epsilon_{2},\epsilon_{1}-\omega+\epsilon_{3}\right)-\bar{\kappa}_{\alpha}\left(\epsilon_{1},\epsilon_{2}+\omega-\epsilon_{3}\right)\right].
\end{eqnarray*}

\twocolumngrid

\bibliographystyle{apsrev4-1}
\bibliography{library}

\end{document}